\documentclass[twocolumn,showpacs,preprintnumbers,amsmath,amssymb,prl,epsf,superscriptaddress]{revtex4}
\usepackage{graphicx}
\usepackage{nicefrac}
\usepackage[squaren]{SIunits}

\begin{document}
\title{Generation and Direct Detection of Broadband Mesoscopic \\Polarization-Squeezed Vacuum}
\author{Timur Iskhakov}
\affiliation{Department of Physics, M.V.Lomonosov Moscow State University, \\ Leninskie Gory, 119992 Moscow, Russia}
\author{Maria~V.~Chekhova}
\affiliation{Max Planck Institute for the Science of Light, G\"unther-Scharowsky-Stra\ss{}e 1/Bau 24, 91058 Erlangen, Germany}
\affiliation{Department of Physics, M.V.Lomonosov Moscow State University, \\ Leninskie Gory, 119992 Moscow, Russia}
\author{Gerd Leuchs}
\affiliation{Max Planck Institute for the Science of Light, G\"unther-Scharowsky-Stra\ss{}e 1/Bau 24, 91058 Erlangen, Germany}
\affiliation{University Erlangen-N\"urnberg, Staudtstrasse 7/B2, 91058 Erlangen, Germany}
\begin{abstract}
Using a traveling-wave OPA with two orthogonally oriented type-I BBO crystals pumped by picosecond pulses, we generate vertically and horizontally polarized squeezed vacuum states within a broad range of wavelengths and angles. Depending on the phase between these states, fluctuations in one or another Stokes parameters are suppressed below the shot-noise limit. Due to the large number of photon pairs produced, no local oscillator is required, and 3dB squeezing is observed by means of direct detection.

\end{abstract}
 \maketitle \narrowtext
\vspace{-10mm}

Squeezed light is the basic resource for continuous-variable quantum communication and computation~\cite{Bachor}. Nowadays, the techniques of generating squeezed light are well developed and include $\chi^{(3)}$ interactions in fibres~\cite{Kerr} and $\chi^{(2)}$ - based optical parametric amplifiers (OPAs), seeded~\cite{Aytur,seeded OPA} or not seeded~\cite{unseeded OPA}, cavity-based~\cite{cavity} or single-pass~\cite{single-pass}. Squeezed light, as a rule, is detected through homodyne measurement, using a local oscillator, or, in the case of bright two-mode squeezing, also by means of direct detection. More advanced techniques of measuring squeezed light include Wigner-function tomography~\cite{WFtomo} and polarization tomography~\cite{Karmas,Marquardt}.

Among the variety of squeezed states, a special role belongs to the squeezed vacuum, a state generated at the output of a parametric down-converter with a vacuum input state~\cite{MandelWolf}. In the limit of weak parametric gain, squeezed vacuum is known as biphoton light~\cite{DNK}, while in the strong-gain limit it manifests quadrature squeezing~\cite{MandelWolf}. At any value of the parametric gain, squeezed vacuum contains only even-photon-number Fock states and is therefore always nonclassical. Initially such states of light were therefore called two-photon coherent states~\cite{Yuen}. Thus, squeezed vacuum plays an important role in the quantum optics of both discrete and continuous variables and provides a useful resource for quantum information protocols.

In particular, two-mode squeezed vacuum generated at the output of an unseeded two-mode parametric downconverter~\cite{Braunstein} has such an important advantage as perfect two-mode squeezing: regardless of the parametric gain, the photon-number difference for the two modes does not fluctuate. Two-mode squeezed vacuum generated in a single-pass OPA is especially interesting, first, because its correlations are not reduced by the cavity losses and second, because it has a rich broadband (multimode) frequency and angular spectrum, which can be used for quantum information protocols in higher-dimensional Hilbert spaces. However, because of the difficulties in producing and detecting such states, there are only a few reports on the direct detection of broadband squeezed vacuum, and all of them show only a small degree of squeezing~\cite{Lugiato,Bondani,Brida,JETPLett}.

In this paper, we achieve a considerable degree of two-mode broadband squeezing by combining two coherent strongly pumped single-pass type-I OPAs, which generate, via parametric down-conversion (PDC), squeezed vacuums in two orthogonal polarization modes. This creates a squeezed vacuum with the variance of one of the Stokes operators suppressed below the shot-noise level, which in this case is given by the mean number of photons~\cite{hidden}. This special type of two-mode squeezing is often referred to as polarization squeezing~\cite{hidden}, and the state can be called polarization-squeezed vacuum. The state is measured by means of direct detection.

\begin{figure}
\includegraphics[width=0.4\textwidth]{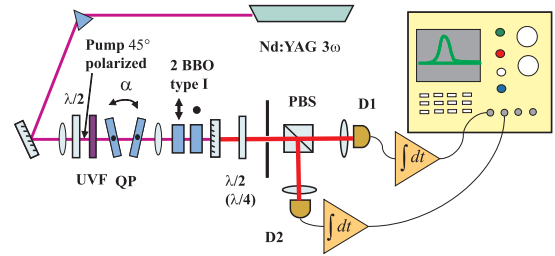}
\caption{The experimental setup.} \end{figure}

In our setup (Fig.1), two  1-mm BBO crystals with the optic axes oriented in orthogonal (vertical and horizontal) planes were placed into the beam of a Nd:YAG laser third harmonic (wavelength $355$nm). The fundamental and second-harmonic radiation of the laser was eliminated with the help of a prism and a UV filter. The pump pulse width was 17 ps, the repetition rate 1 kHz, and the mean power up to 120 mW. The pump was focused into the crystals by either a lens with focal length 100 cm, which resulted in a beam waist of $70 \mu$m, or with a telescope, providing a more soft focusing (beam waist about $500\mu$m). Using a half-wave plate, we aligned the pump polarization to be at $45^{\circ}$ to the planes of the crystals optic axes, so that the pump provided equal contributions to PDC in both crystals. The crystals were cut for PDC with type-I collinear frequency-degenerate phase matching. After the crystals, the pump radiation was cut off by two dichroic mirrors, having high reflection for the pump and $98.5\%$ transmission for PDC radiation. The detection part of the setup, including a polarizing beam splitter, a half- or quarter-wave plate and two detectors, provided a standard Stokes measurement. All surfaces of the optic elements had a standard broadband antireflection coating. The angular spectrum of the detected light was restricted by an aperture to be $0.8^{\circ}$, and the wavelength range was only restricted by PDC phase matching and was $130$ nm broad, with the central wavelength $709$nm. The detectors, Hamamatsu S3883 p-i-n diodes followed by pulsed charge-sensitive amplifiers~\cite{Hansen} based on Amptek  A250 and A275 chips, with peaking time \unit{2.77}{\micro\second}, had a quantum efficiency of $90\%$ and electronic noise equivalent to $180$ electrons/pulse. At their outputs, they produced pulses of a given shape, with the duration \unit{8}{\micro\second} and the amplitude proportional to the integral number of photons per light pulse. The phase between the states generated in the two crystals could be varied by tilting two quartz plates with thicknesses 532 and 523 $\mu$m, placed into the pump beam and having the optic axes oriented vertically.

\begin{figure}
\includegraphics[width=0.2\textwidth]{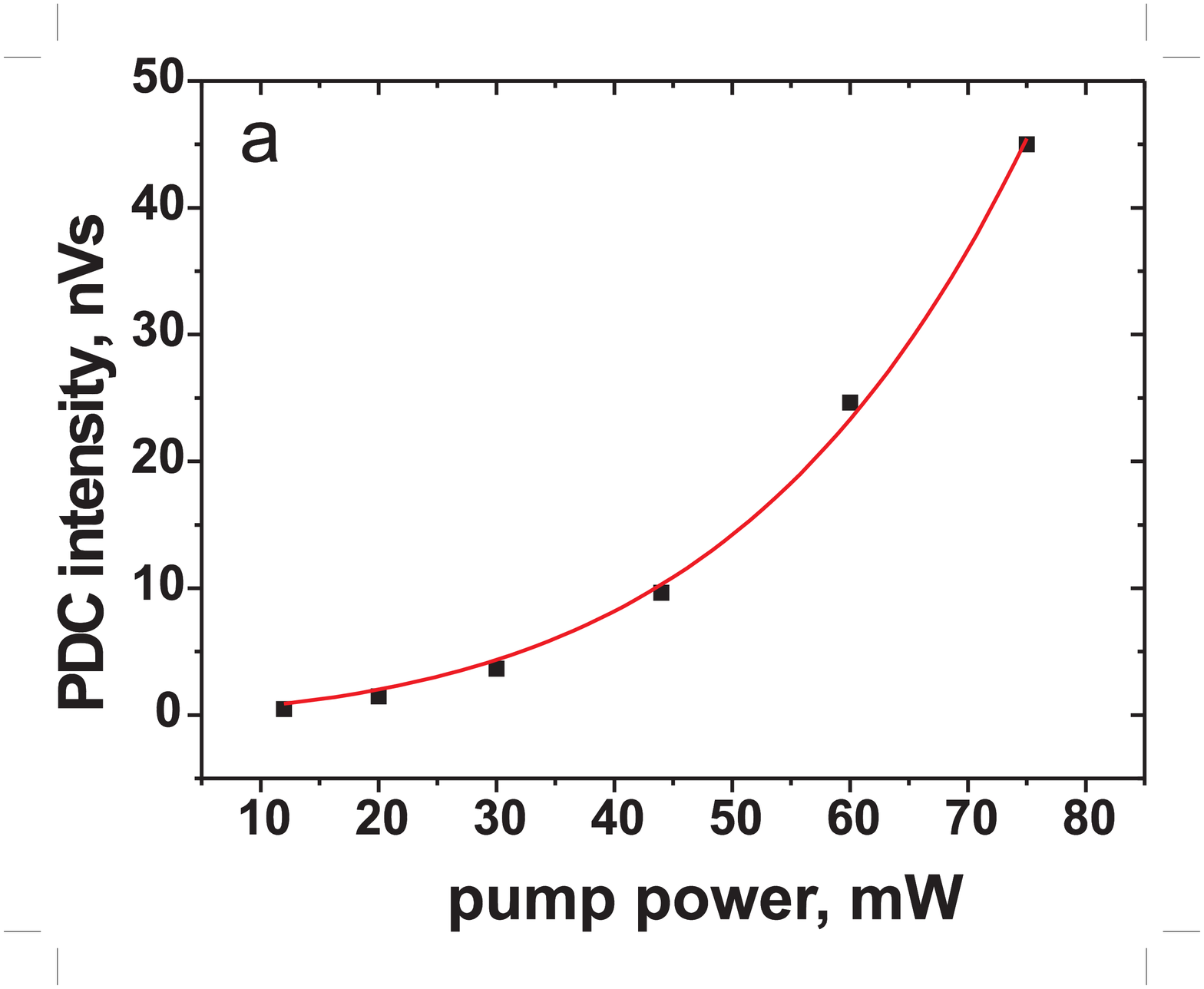}
\includegraphics[width=0.207\textwidth]{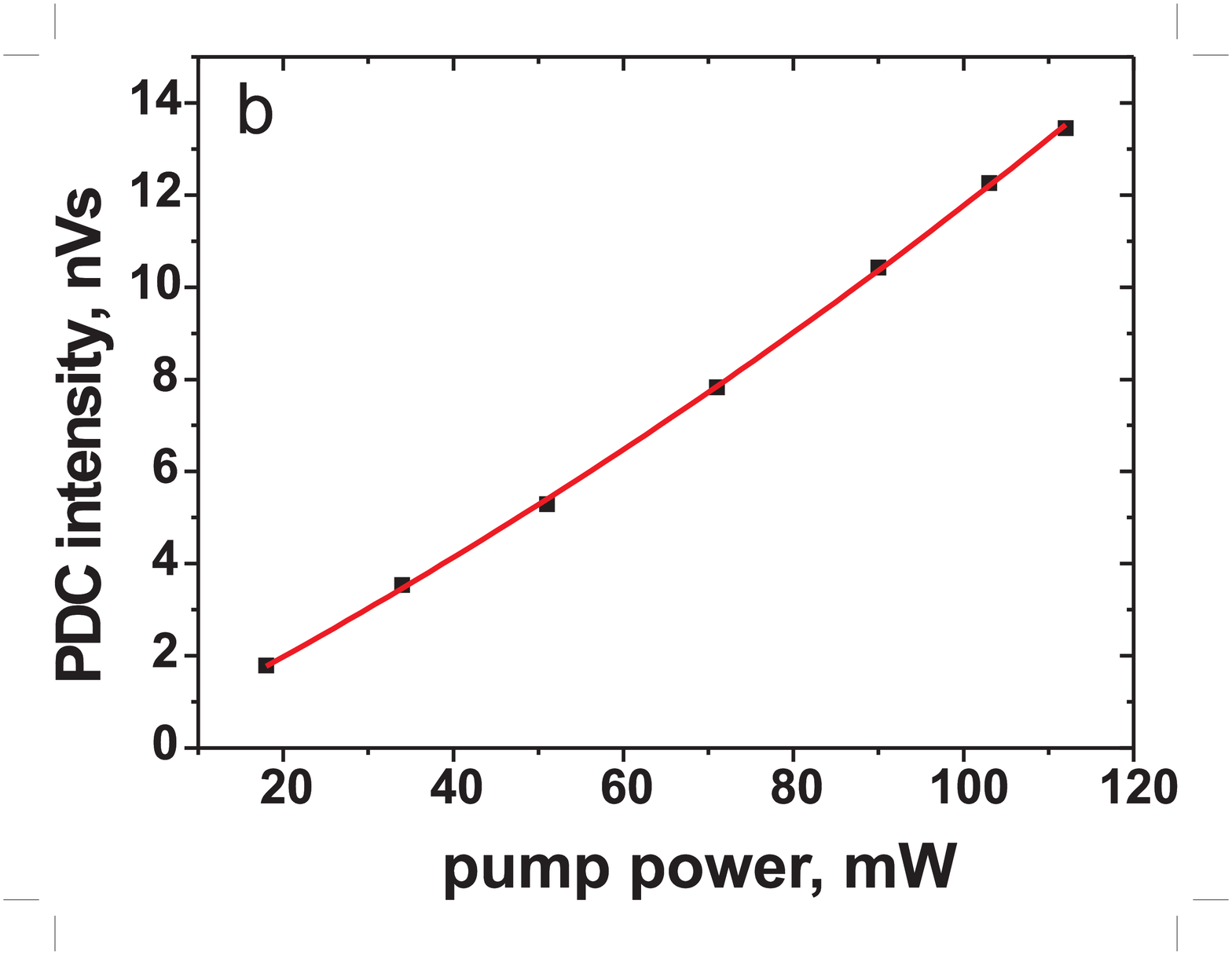}
\caption{Dependence of the OPA output on the pump power (a) for \unit{70}{\micro\meter} beam waist and (b) for \unit{500}{\micro\meter} beam waist. Solid lines show a fit with Eq.(\ref{sinh}).} \end{figure}

The output signals of the detectors were measured by means of an AD card integrating the electronic pulses over time. The result coincided, up to the amplification factor $A$, with the photon numbers incident on the detectors during a light pulse. The amplification factors for detectors 1 and 2 were independently calibrated to be $A_1=\unit{9.96\cdot 10^{-3}}{\nano\volt\second}$/photon and $A_2=\unit{1.107\cdot 10^{-2}}{\nano\volt\second}$/photon. The difference between the detectors' amplification factors was eliminated numerically, by multiplying the result of the measurement for detector 2 by a factor of $0.9-0.92$, depending on the alignment. As a result, the output signals of the detectors were balanced to an accuracy of $0.1\%$. From the dataset obtained for 30000 pulses, mean photon numbers per pulse were measured, as well as the variances of the photon-number difference and photon-number sum for the two detectors. Since the electronic noise was not small compared to the shot-noise level, it had to be subtracted. The shot-noise level was measured independently, by using a shot-noise limited laser source, and was on the order of several hundreds of photons per pulse. The degree of two-mode squeezing was characterized~\cite{Aytur,Lugiato} by the noise reduction factor (NRF) defined as the ratio of the photon-number difference variance to the mean photon-number sum,

\begin{equation}
\hbox{NRF}=\frac{Var(N_1-N_2)}{\langle N_1+N_2\rangle}.
\label{NRF}
\end{equation}

Dependence of the detector output signal on the input pump power $P$ is shown in Fig.2. The measurement was made for a 10 nm wavelength band selected by means of an interference filter, and the collection angle of $0.4^{\circ}$ selected by an aperture. With the pump focused tightly (using a lens with focal length 100 cm), the dependence is strongly nonlinear (Fig.2a), and by fitting it with the well-known expression for a single-pass OPA output~\cite{QE},

\begin{equation}
N=m\sinh^2\Gamma,
\label{sinh}
\end{equation}
with $m$ denoting the number of modes and the parametric gain scaling as square root of the pump power, $\Gamma=\kappa \sqrt{P}$, $\kappa$ and $m$ being the fitting parameters, we estimate the highest parametric gain achieved in our measurement as $\Gamma=3.4$. With soft focusing, by means of a telescope, the $N(P)$ dependence is much less nonlinear (Fig.2b), and the highest gain achieved, according to the fit, is only $0.8$. Note that the dependences shown in Figs. 2a,b were obtained with the pump polarized vertically, so that only one crystal contributed into the signal; with the pump polarized at $45\deg$, both crystals contributed but the gain $\Gamma$ in each of them was reduced $\sqrt{2}$ times.
From the value of the gain we see that the number of photons per mode ranges between one and a hundred, i.e., we are in the regime of mesoscopic twin beams~\cite{Perina}. At the same time, the largest total number of photons per pulse is about 4000 for tight focusing and about 1500 for soft focusing. Without the interference filter, photon numbers per pulse are about one order of magnitude higher.

\begin{figure}
\includegraphics[width=0.25\textwidth]{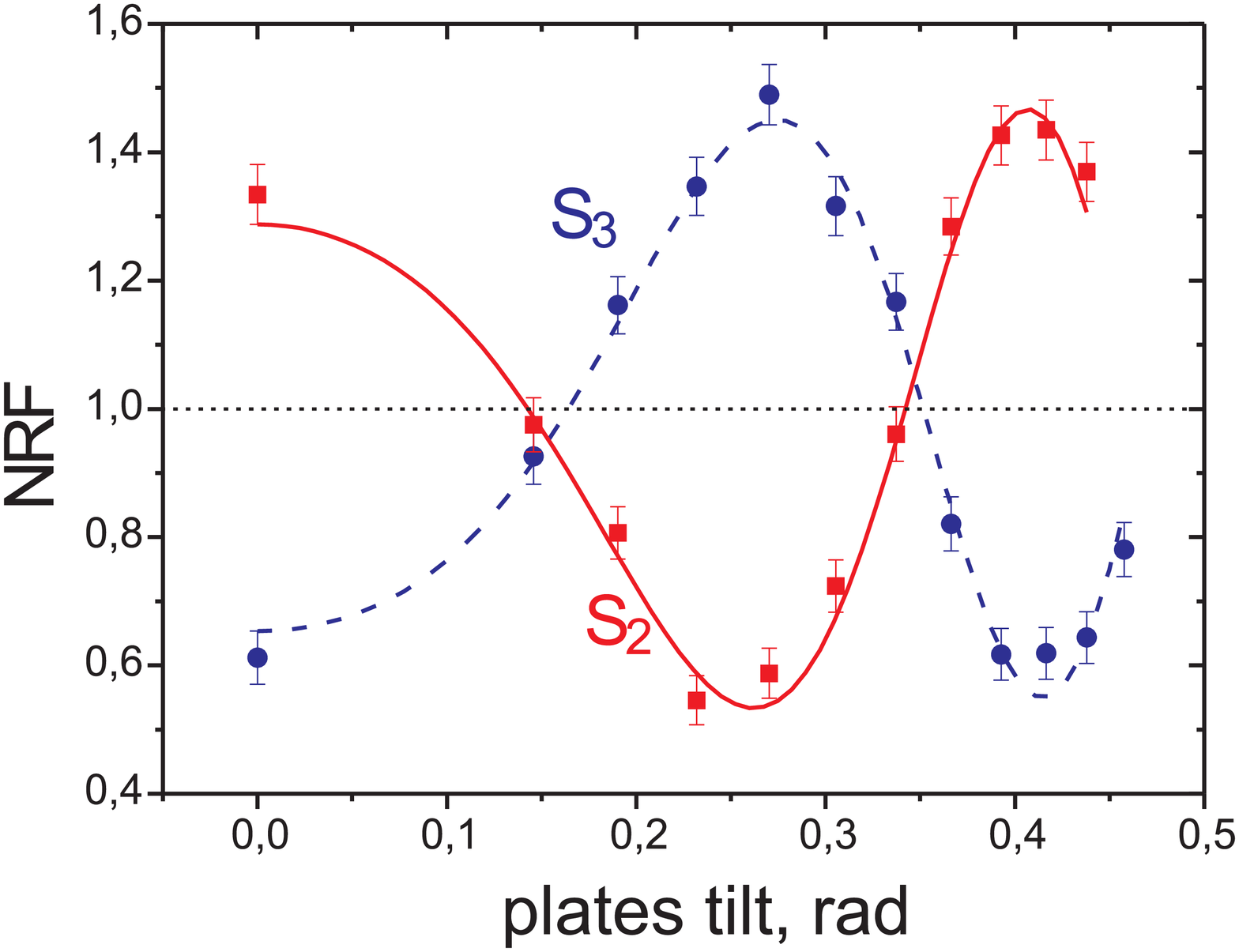}
\caption{Noise reduction factor in $S_2$ (squares, solid line) and $S_3$ (circles, dashed line) depending on the quartz plates tilt. } \end{figure}

Figure 3 shows the the variances of the second and the third Stokes operators measured versus the delay introduced by the two quartz plates between the vertically and horizontally polarized components of the pump beam. The measurement was performed at low parametric gain, with the $80$ mW pump beam softly focused into the crystals. When the phase delay was equal to $\pi$, squeezing was obtained in the $S_2$ Stokes observable, while $S_3$ was anti-squeezed. When the phase delay was equal to zero, the squeezed Stokes observable was $S_3$. At the same time, the observable $S_1$ was always anti-squeezed. This behavior is clear from the following simple considerations: the states generated at the outputs of the two crystals consist of horizontally and vertically polarized photon pairs, $|HH\rangle$ and $|VV\rangle$. Due to quantum interference, they become orthogonally polarized pairs. In particular, the sum $|HH\rangle+|VV\rangle$ gives two pairs of right- and left-circularly polarized photons, $|RL\rangle$, while the difference, $|HH\rangle-|VV\rangle$, gives two pairs of diagonally and anti-diagonally polarized photons, $|AD\rangle$. Therefore, the $\pi$ phase delay between the squeezed vacuum fields generated in the two crystals will lead to a state with suppressed fluctuations in the second Stokes operator, while for the case of the zero phase delay, the state will be squeezed in the third Stokes operator. Note that the mean values of all Stokes observables $\langle S_1\rangle,\langle S_2\rangle,\langle S_3\rangle$ are equal to zero; the state is unpolarized in the 'classical optics' sense but polarized in second- and higher-order moments. This property is called in literature 'hidden polarization'~\cite{hidden,Chirkin,DNK92} and led to the introduction of various definitions for the degree of polarization~\cite{Chirkin, dp}.

A rigorous calculation can be made in the framework of the Heisenberg approach. Denoting the photon creation operators in the horizontal and vertical polarization modes as $a_h^{\dagger}$ and $a_v^{\dagger}$, we can write the Hamiltonian of the OPA as the sum of the Hamiltonians corresponding to separate crystals,

\begin{equation}
\hat{H}=G[(a_h^{\dagger})^2+e^{i\phi}(a_v^{\dagger})^2]+\hbox{h.c.},
\label{Hamiltonian}
\end{equation}
where $G$ is proportional to the pump field amplitude and $\phi$ is the phase delay introduced between the vertical and horizontal pump components. Then the Hamiltonian (\ref{Hamiltonian}) can be expressed via photon creation operators in orthogonal elliptically polarized modes with the axes of the ellipses oriented at $\pm 45\deg$,  $a_{\phi}^{\dagger}\equiv1/\sqrt{2}(a_h^{\dagger}+ie^{i\phi/2}a_v^{\dagger})$ and $b_{\phi}^{\dagger}\equiv1/\sqrt{2}(a_h^{\dagger}-ie^{i\phi/2}a_v^{\dagger})$~\cite{Burlakov}:

\begin{equation}
\hat{H}=2Ga_{\phi}^{\dagger}b_{\phi}^{\dagger}+\hbox{h.c.}
\label{Ham1}
\end{equation}

From the form of the Hamiltonian it is clear that the output state has perfectly correlated fluctuations in modes corresponding to the operators $a_{\phi}, b_{\phi}$. The mean values and variances of all Stokes operators can be calculated using the Bogoliubov transformations following from the Hamiltonian (\ref{Ham1}):

\begin{equation}
a_{\phi}=U a_{0\phi}+V b_{0\phi}^{\dagger},\,\,b_{\phi}=U b_{0\phi}+V a_{0\phi}^{\dagger},
\label{Bogoliubov}
\end{equation}
where $U=\cosh\Gamma$ and $V=\sinh\Gamma$, $\Gamma$ is dimensionless gain proportional to $G$, and $a_{0\phi}, b_{0\phi}$ are the input (vacuum) photon annihilation operators. Losses in the optical elements and non-unity quantum efficiencies of the detectors can be taken into account in the usual way, by introducing a beamsplitter in front of each detector, with the amplitude transmission coefficient $\sqrt{\eta}$ and the amplitude reflection coefficient $\sqrt{1-\eta}$.

Calculation of the variances of the Stokes operators $\hat{S}_2\equiv a_h^{\dagger}a_v+a_v^{\dagger}a_h$ and $\hat{S}_3\equiv -i(a_h^{\dagger}a_v-a_v^{\dagger}a_h)$ is performed by passing to $a_{\phi}, b_{\phi}$ operators, then applying transformations (\ref{Bogoliubov}), and then averaging the corresponding second-order moments over the vacuum state. The result, with an account for losses $\eta$, is

\begin{eqnarray}
\frac{\hbox{Var}(S_2)}{2\eta V^2}=(1-\eta)\sin^2\frac{\phi}{2}+(2\eta U^2+1-\eta)\cos^2\frac{\phi}{2},\nonumber\\
\frac{\hbox{Var}(S_3)}{2\eta V^2}=(1-\eta)\cos^2\frac{\phi}{2}+(2\eta U^2+1-\eta)\sin^2\frac{\phi}{2}.
\label{Var}
\end{eqnarray}

Because the total number of registered photons is given by the expression $\langle S_0\rangle=2\eta V^2$, the left-hand sides of Eqs.(\ref{Var}), according to Eq.(\ref{NRF}), give the NRF values for the $\hat{S}_2$ and $\hat{S}_3$ Stokes operators. In the low-gain limit (as is the case for the above-described measurement), $U\approx1$, and Eqs. (\ref{Var}) become

\begin{equation}
\hbox{NRF}(S_2)=1+\eta\cos\phi, \, \hbox{NRF}(S_3)=1-\eta\cos\phi
\label{lg}
\end{equation}

Equations (\ref{lg}) were used to fit the data shown in Fig.3, taking into account the relation between the tilt angle of the plates $\alpha$ and the phase $\phi$. The only fitting parameters being the efficiency $\eta$ and the initial phase delay, the theoretical dependence is in a good agreement with the experimental data. The resulting quantum efficiency is found to be $\eta=0.45$, which is much less than we expect from the values of the detectors' quantum efficiencies and the optical losses (about 7\%).

However, in our experimental scheme there is one more source of losses. Because the nonlinear crystals are placed into the pump beam one after another, squeezed vacuum generated in the first crystal passes through the second crystal. As it was shown in Ref.~\cite{spectra} for the case of two-photon light, the desired state is only produced in the central part of the frequency-angular spectrum. The state produced at the 'slopes', due to the group-velocity dispersion and optical anisotropy, is a different one. In our case the angular bandwidth is restricted by the aperture, but the frequency spectrum contains the whole band allowed by phasematching. If the crystals are aligned to produce, for instance, $S_2$-squeezed state at the central wavelength (709 nm), the state generated at the 'slopes' will be $S_3$ - squeezed and anti-squeezed in $S_2$. According to our numerical calculation, the fraction of the 'correct' state for the case of degenerate phase matching is $0.71$. This explains the low degree of noise suppression in our experiment.

\begin{figure}
\includegraphics[width=0.27\textwidth]{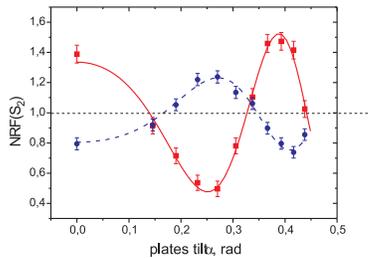}
\caption{Noise reduction factor in $S_2$ depending on the quartz plates tilt for the degenerate (squares) and non-degenerate (circles) phase-matching.} \end{figure}

The effect of group-velocity dispersion in the second crystal reduces the degree of squeezing even more drastically in the case of non-degenerate phase matching. For instance, for the phase matching at $650$ nm and $780$ nm, numerical calculation shows that the fraction of the squeezed state in the bandwidth is only $0.52$. To test this in experiment, we measured the $S_2$ variance for two cases: the crystals aligned for degenerate phase matching and the crystals aligned for phasematching at wavelengths $650$ nm and $780$ nm (Fig.4). While in the first case the NRF reaches values below $0.5$, in the second one the minimal value of NRF is $0.75$. Note that the phase of the interference dependence is also changed in the degenerate regime, because of the additional tilt of the crystals.

We would like to stress that this effect is not inevitable and is only present in configurations where two crystals are placed one after another. For instance, a setup with a Mach-Zehnder interferometer~\cite{Burlakov} is free from this drawback. Here, we used the construction with two crystals in a single beam only for the sake of higher stability; however, if the goal is to achieve high degree of squeezing, a Mach-Zehnder interferometer is a much better choice. Note that there is a recent proposal on a similar scheme employing a high-stability Mach-Zehnder interferometer based on birefringent beam displacers~\cite{fiorentino}.

In conclusion, we report a source of broadband twin beams (squeezed vacuum) covering the wavelength range from 650 to 780 nm and the angular range up to $0.8^{\circ}$. The source is a mesoscopic one, providing the number of photons per mode between 1 and 100. Because of its highly multimode character, the resulting number of photons per pulse is up to $5\cdot 10^5$. Polarization properties of the produced state reveal 'hidden polarization' effect: depending on the polarization of the pump beam, the output state has fluctuations in either $S_2$ or $S_3$ Stokes observables squeezed 50\% below the shot-noise level. To the best of our knowledge, this is the first report on such a high degree of squeezing observed via direct detection of twin beams from a single-pass OPA.

We acknowledge the financial support of the European Union under project COMPAS No. 212008 (FP7-ICT). M.V.C. acknowledges the support of the DFG foundation.
We are grateful to Alan Huber from Amptek company for his valuable advice on constructing low-noise charge-integrating detectors and to Bruno Menegozzi for the realization of the electronic circuit.


\begin{references}

\bibitem{Bachor} H.-A.~Bachor, T.~C.~Ralph, \emph{A Guide to Experiments in Quantum Optics}
(Wiley-VCH, sec. ed., Weinheim, 2004).

\bibitem{Kerr} R.M. Shelby et al., Phys. Rev. Lett., \textbf{57}, 691 (1986); J.~Heersink et al., Phys. Rev. A \textbf{68}, 013815 (2003); J.~Heersink et al., Opt. Lett. \textbf{30}, 1192 (2005).

\bibitem{Aytur}O. Aytur, P. Kumar, Phys. Rev. Lett. \textbf{65}, 1551 (1990).

\bibitem{seeded OPA}  D.T. Smithey et al., Phys. Rev. Lett. \textbf{69}, 2650 (1992).

\bibitem{unseeded OPA} A. Heidmann et al., Phys. Rev. Lett. \textbf{59}, 2555 (1987).

\bibitem{cavity}Ling-An Wu et al., Phys. Rev. Lett. \textbf{57}, 2520 (1986); J.~Gao et al., Opt. Lett. \textbf{23}, 870 (1998).

\bibitem{single-pass} J.~A.~Levenson et al., Quantum Semiclass. Opt. \textbf{9}, 221 (1997); T.~Hirano et al., Opt. Lett. \textbf{30}, 1722 (2005).

\bibitem{WFtomo} K.~Vogel and H.~Risken, Phys. Rev. A \textbf{40}, 2847 (1989); D.T. Smithey et al., Phys. Rev. Lett. \textbf{70}, 1244 (1993).

\bibitem{Karmas} V.~P.~Karassiov and A.~V.~Masalov, J. Opt. B: Quantum Semiclass. Opt. \textbf{4}, S366 (2002); P.~A.~Bushev et al., Optics and Spectroscopy \textbf{91}, 526 (2001).

\bibitem{Marquardt} Ch.~Marquardt et al., Phys. Rev. Lett. \textbf{99}, 220401 (2007).

\bibitem{MandelWolf} L.~Mandel and E.~Wolf, \emph{Optical Coherence and Quantum Optics} (Cambridge University Press, Cambridge, 1995).

\bibitem{DNK} D.N.~Klyshko, \emph{Photons and Nonlinear Optics} (Gordon and
Breach, New York, 1988).

\bibitem{Yuen} H.P. Yuen, Phys.Rev. A \textbf{13}, 2226 (1976).

\bibitem{Braunstein} S.~L.~Braunstein, Phys. Rev. A \textbf{71}, 055801 (2005).

\bibitem{Lugiato} O. Jedrkiewicz et al., Phys. Rev. Lett. \textbf{93}, 243601 (2004).

\bibitem{Bondani} M. Bondani et al., Phys. Rev. A \textbf{76}, 013833 (2007).

\bibitem{Brida} G.~Brida et al., Journal of Modern Optics, to appear (2008).

\bibitem{JETPLett} T.~Sh.~Iskhakov et al., JETP Lett., to appear (2008).

\bibitem{hidden} V. P. Karasev, A.V. Masalov, Opt. Spectrosc. 74, 551 (1993); P. Usachev et al. Opt. Commun. 193, 161 (2001).

\bibitem{Hansen} H.~Hansen et al.,
Optics Letters \textbf{26}, 1714 (2001).

\bibitem{QE}O.~A.~Ivanova et al., Quant. Electron. \textbf{36} (10), 951 (2006).

\bibitem{Perina} J.~Perina et al., Phys. Rev. A \textbf{76}, 043806 (2007).

\bibitem{Chirkin} A. S. Chirkin, A. A. Orlov, D. Yu. Paraschuk, Quantum Electron. 23, 870 (1993); A. P. Alodjants, S. M. Arakelyan, and A. S. Chirkin, JETP \textbf{81}, 34 (1995).

\bibitem{DNK92} D.~N.~Klyshko, Physics Letters A \textbf{163} 349, (1992).

\bibitem{dp} V. P. Karassiov, J. Phys. A \textbf{26}, 4345 (1993); D.~N.~Klyshko, JETP \textbf{84}, 1065 (1997); A. Luis, Phys. Rev. A \textbf{66}, 013806 (2002); G.~Bj\"ork et al., Proc. SPIE \textbf{4750} (2002); A. B. Klimov et al., Phys. Rev. A \textbf{72}, 033813 (2005).

\bibitem{Burlakov} A.~V.~Burlakov et al., Phys. Rev. A \textbf{64}, 041803(R) (2001).

\bibitem{spectra} G.~Brida et al., Phys. Rev. A \textbf{76}, 053807 (2007).

\bibitem{fiorentino} M.~Fiorentino and R.G. Beausoleil, Opt. Exp. \textbf{16}, 20149 (2008).

\end{references}
\end{document}